# The TA Framework: Designing Real-time Teaching Augmentation for K-12 Classrooms

Pengcheng An[1,*]   Kenneth Holstein[2,*]   Bernice d'Anjou[1]   Berry Eggen[1]   Saskia Bakker[3]

[1] Department of Industrial Design, Eindhoven University of Technology, Eindhoven, NL
[2] Human-Computer Interaction Institute, Carnegie Mellon University, Pittsburgh, PA, USA
[3] Philips Experience Design, Eindhoven, NL

{p.an, j.h.eggen}@tue.nl;  kjholste@cs.cmu.edu;  b.e.m.d.anjou@student.tue.nl;  saskia.bakker@philips.com

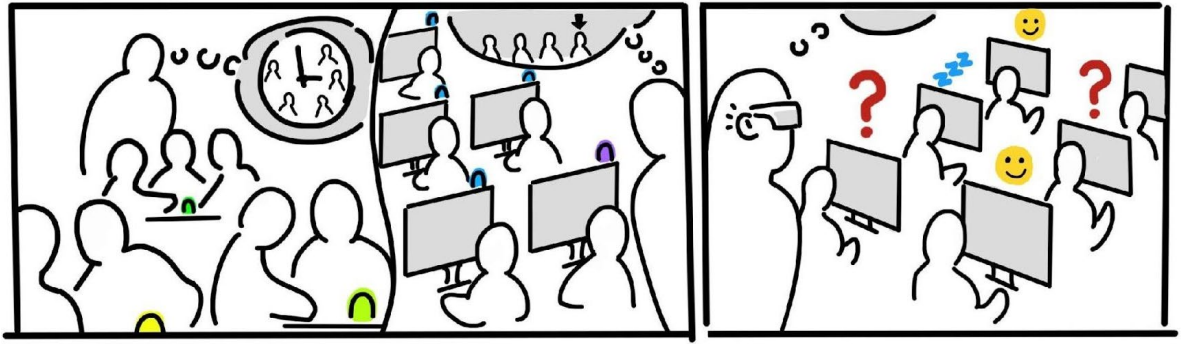

**Figure 1.** Two examples of real-time teaching augmentation. LEFT: ambient information on light objects supporting teachers' (a) division of time and attention over students [10] and (b) monitoring of computer-supported learning processes [12]. RIGHT: mixed reality glasses showing real-time indicators floating above students' heads, based on AI analytics, to direct teachers' attention towards students who may need it most [55].

**ABSTRACT**
Recently, the HCI community has seen increased interest in the design of *teaching augmentation* (TA): tools that extend and complement teachers' pedagogical abilities *during ongoing classroom activities*. Examples of TA systems are emerging across multiple disciplines, taking various forms: e.g., ambient displays, wearables, or learning analytics dashboards. However, these diverse examples have not been analyzed together to derive more fundamental insights into the design of teaching augmentation. Addressing this opportunity, we broadly synthesize existing cases to propose the *TA framework*. Our framework specifies a rich design space in five dimensions, to support the design and analysis of teaching augmentation. We contextualize the framework using existing designs cases, to surface underlying design trade-offs: for example, balancing actionability of presented information with teachers' needs for professional autonomy, or balancing unobtrusiveness with informativeness in the design of TA systems. Applying the TA framework, we identify opportunities for future research and design.

**CSS Concepts**
• **Human-centered computing~Interaction design~ Interaction design theory, concepts and paradigms**

**INTRODUCTION**
Classroom teaching is a complex job: teachers must frequently improvise and constantly seek the best fit between their actions and the needs of their students [26, 117]. For instance, while teaching, a teacher has to continuously observe each learner's current state, in order to reflectively decide which learners are in need of support and what types of support should be offered [65, 124]. With an increased emphasis on personalized learning [96], teachers are now tasked with even greater challenges in tracking students' learning processes and adapting accordingly [37, 97].

Given these challenges, recent years have seen increased interest in the design of technologies to augment teachers' practice during ongoing classroom activities. At the intersection of HCI and education, research has predominantly focused on the design of technologies to support *learners* (e.g., [3, 8, 43, 75, 78, 121]). Nonetheless, a sizeable and fast-growing body of literature, spanning multiple disciplines, has focused on supporting *teachers* in the midst of their day-to-day practice, i.e., while they are teaching (e.g., [10, 12, 55, 66, 88]).

In this paper, we frame these diverse technologies broadly as real-time *teaching augmentation* (TA). In the spirit of prior HCI work on intelligence augmentation (e.g., [35, 73, 114]), we define TA systems as technologies that extend and complement teachers' pedagogical capabilities *in action* (e.g., informing teachers' reflection and decision-making by providing relevant and timely information during an ongoing lesson [10, 50]; see Figure 1).

Although a range of TA systems have been developed, these diverse cases have not been analyzed together to generalize more fundamental, *intermediate-level design knowledge* [57]: i.e., design knowledge with applicability across different types of teaching augmentation systems. For example, some TA systems show information publicly in the classroom [4, 12], while others present data only to teachers [55, 103]. Some TA systems enhance teaching via low-resolution, glanceable information [12, 55] while others offer rich, focal visualizations [82, 127]. While each solution naturally has its advantages and disadvantages, these have not been clearly laid out by analyzing multiple designs together. Such (and many other kinds of) intermediate-level knowledge is still scarce in educational technology research [18, 79, 99]. Yet the communication of such knowledge has proven fruitful in various HCI domains [56, 83, 130, 133]: for instance, in helping designers and researchers to explicate crucial yet implicit design tensions or trade-offs, or explore relevant but less-charted regions of the design space.

This paper sets out to generate intermediate-level knowledge relevant to TA systems, by proposing a framework for the design and analysis of teaching augmentation: the TA framework. To develop this framework, we conducted a broad synthesis of prior TA design cases, through a constant comparative analysis [46]. The framework describes a design space in five dimensions, representing important design choices that have received relatively little discussion in prior literature: *augmentation target*, *attention*, *social visibility*, *presence over time*, and *interpretation*. The TA framework can help designers and researchers describe and analyze different designs to surface new insights and possibilities. In this paper, we illustrate and contextualize the five dimensions of the TA framework using design cases from prior literature. We then demonstrate the framework's utility by analyzing patterns in existing TA designs, which we use to identify opportunities for future exploration.

In summary, this paper contributes the TA framework, which aims to inform the design and analysis of teaching augmentation systems by: (1) revealing underlying, often implicit design considerations with relevance across diverse kinds of TA systems, and (2) proposing a common lens through which researchers in this interdisciplinary area can analyze existing design cases and generate new insights.

## RELATED WORK ON TEACHING AUGMENTATION

TA systems have been developed across a range of disciplines, including HCI [11, 12, 66], learning analytics [83, 86, 127], teacher education [58, 111, 112], AI in education [40, 55, 125, 141], and the learning sciences [71, 122]. Their designs take various forms (e.g., see Figure 4), such as dashboard interfaces (e.g., [88, 90, 92, 126]) peripheral information displays (e.g., [4, 7, 13, 69, 89]), or wearables (e.g., [19, 55, 58, 103]). These systems have been designed to provide support for a range of classroom contexts, from primary [13, 129] to secondary [12, 55] to higher [4, 81] education, and from traditional [129] to blended [52] classrooms. Furthermore, TA tools have been designed to support a range of teacher processes, many of which are covered by the notion of *classroom orchestration* [30, 71, 125], such as monitoring [61, 107, 118, 122] and decision-making [12, 90, 92, 135]. TA tools have also been designed for teachers' professional development [37, 93], such as real-time teacher coaching [58, 112], or cultivating their on-the-spot reflectiveness [10, 66]. In this section, we present an overview of existing TA systems. These prior cases motivate the present work and form the basis for our generalization of the TA framework.

Real-time *learning analytics dashboards* [88, 90, 118, 127] are commonly designed to extend teachers' in-the-moment awareness in blended classrooms by presenting teachers with real-time information about students' progress and performance as they work with educational software [51, 83, 86, 122, 125]. This information is often displayed via lists, tables, or data visualizations (e.g., line plots and bar charts) on PCs or handheld mobile devices [2, 25, 41]. Some of these systems include broader support for *classroom orchestration* [30]; e.g., by helping teachers dynamically transition students between individual and group activities, or enabling them to discretely communicate with students during class via a tablet interface [30, 52, 94, 125].

The HCI notion of *calm technology* [137] argues that computing systems should be able to engage users in the periphery of their attention instead of only demanding their focal attention. Inspired by this notion, a number of TA systems have been designed as *public peripheral (ambient) information displays* [69, 84, 98], to extend teachers' (as well as students') situational awareness in a class session. Lernanto [7], for example, uses wall-mounted, arrayed LEDs to depict each student's learning pace [6], which provides teachers with background awareness that can help them in differentiating instruction accordingly. ClassSearch [89] uses a projection to show the ongoing web search behaviors of learners, which can help teachers notice search behaviors that they can comment on to give helpful instructions (e.g., proper use of quotation marks) [89]. CawClock [15] uses both visuals and soundscapes to afford teachers' and pupils' awareness of different activity sessions planned in a lesson.

Reflecting a trend towards *ubiquitous computing* [136] in the classroom, several TA systems use *distributed* digital lamps [4, 11–13, 31, 129] to convey ambient information throughout the physical classroom space. Such distributed systems augment teachers' physical surroundings and situate information in its relevant location [59, 129]. Lantern [4], a classroom orchestration system, uses ambient lamps placed at student desks to display university student-teams' work progress and help-requests via low-resolution signals such as color and pulse rate. Similarly, the FireFlies systems [11–13, 129] (two examples of which are reviewed in the next section) support teachers by opening up non-verbal communication channels between the teacher and students via the color of each student/team's light object.

Related research has also begun to explore teaching augmentation through *wearables*. For example, Quintana et al. explored the use of smartwatches to support a variety of teacher tasks [103]. Moreover, a strand of research in the area of *teacher education* [27] has focused on the design and evaluation of "synchronous coaching" methods in which a coach, who monitors lessons remotely, provides novice teachers with live feedback and advice during unfolding class sessions through earpieces [58, 111, 112]. A few recent projects have explored the use of smart glasses to augment teacher perceptions and decision-making while still keeping their heads up and their eyes focused on the classroom [19, 50, 55, 143]. For example, GlassClass [19] uses Google Glass to show teachers how well the class understands a lecture (via live student ratings), as well as their work progress and help requests. Lumilo [55] (reviewed in the next section) uses mixed reality glasses that superimpose distributed information onto the teachers' environment (e.g., as visual icons floating over students heads; see Figure 1) to support them in orchestrating AI-supported class sessions.

Despite a surge of interest in real-time teaching augmentation across the aforementioned disciplines, these diverse design solutions have not previously been examined together to generalize more fundamental design insights [57, 79, 99]. Prior work has introduced process-oriented frameworks to guide the iterative design of particular kinds of TA systems [53, 83, 86]. Other work has proposed design implications or conceptual frameworks for smart classrooms [44, 128], orchestration systems [31, 100], learning analytics dashboards [126], Open Learner Models [23], or reflection-aiding tools [10, 66, 120]. Yet little work has taken a broad view of teaching augmentation, to map common design considerations across interfaces and applications. This suggests an opportunity to synthesize intermediate-level knowledge [57] for the design of TA. By generalizing the TA framework from diverse prior design cases, we aim to offer a common ground for researchers to design, analyze, and communicate about TA systems emerging from different domains, and thereby to jointly advance future design.

## TWO RECENT DESIGN CASES

In this section, we review two recent TA design cases in greater depth—FireFlies [10, 12] and Lumilo [51, 55]—two *Research through Design* [144] projects independently conducted in Europe and the US. In so doing, we intend to: (1) concretely illustrate how TA systems can augment classroom teaching in real-time, and (2) refer back to these two cases in the following sections, along with others, to contextualize the TA framework through examples. These two projects span diverse settings, including traditional face-to-face [10], blended [12], and AI-supported [55] classrooms. Moreover, these two projects represent rare examples of TA system development that took *design-oriented* approaches to conducting and reporting research [38, 39, 144]. In other words, these two projects include rich documentation about their context of study (specific teacher and student needs), design rationales (how the designs address these needs) and design evaluations (how the designs were experienced in actual classrooms).

**FireFlies: Designing Peripheral Teaching Augmentation to Enhance Teachers' Sensemaking in Routines**

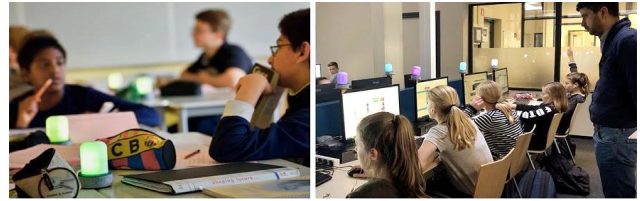

**Figure 2. Two implementations of FireFlies platform** [129]**: (Left) 'ClassBeacons'** [10] **and (Right) 'FireFlies-VLE'** [12]**.**

FireFlies [13, 129] is an open-ended platform to support teachers through distributed wireless lamps that can be pre-programmed for different applications. We review two of its implementations: 'ClassBeacons' [11], and 'FireFlies-VLE' [12]. These two designs were informed by a *context mapping* [119] study for understanding teachers' classroom routines [9]. This study revealed the busy and dynamic nature of teachers' classroom routines through vivid examples [9], which suggested teachers' needs for systems that can both offload and enhance their sensemaking on the fly, without interrupting their nomadic workflows.

One implementation of FireFlies, entitled 'ClassBeacons', visualizes *teacher proximity* [63, 80]: how the teacher divides time and attention over students during a class session. ClassBeacons continuously depicts a teacher's proximity distribution based on real-time *positioning* and *orientation* data [11]. Each lamp of ClassBeacons slowly changes color from yellow to green as the teacher spends more time around a student group (Figure 2). Its field deployment [10] showed that the teachers unobtrusively perceived, or just "*happened to notice,*" the information while performing various tasks within the classroom [10]. ClassBeacons was found to enhance the teachers' *reflection-in-action* [10, 117]. For example, it helped teachers responsively plan their upcoming interactions with learners, such as whom to help next and how much time to spend with them. Moreover, teachers appreciated that ClassBeacons was designed as a "neutral" portrayal of their whereabouts using a yellow-green spectrum (instead of including judgmental colors like red). This was because the distribution of teacher proximity in a lesson is highly *context-dependent* [10], making it difficult for a system to assess whether it is 'good' or 'bad'. Therefore, teachers appreciate being able to interpret the display based on their own *contextual knowledge* [10]. The results also suggested that this public display could increase students' *accountability* [11, 36] for teacher proximity.

A second implementation of FireFlies, called FireFlies-VLE (Figure 2), helps teachers monitor lessons in which students use a virtual learning environment (VLE). Each lamp depicts both a student's performance and concentration in the VLE system (based on analytics described in [138]). A purple–blue color spectrum is mapped to students' low or high concentration levels; an insufficient score at the end of an exercise is indicated by a breathing, dimming effect [12]. A field deployment [12] showed that FireFlies-VLE offered an overview of each learner's general state without requiring teachers to closely check student screens, allowing teachers to offer punctual support and make efficient use of class time. Given the mutual awareness fostered by FireFlies-VLE (i.e., students see the same information the teacher sees), students were prompted to open up to the teacher about their difficulties, boredom, or distraction during class. While 95% of students perceived the system positively [12], qualitative data indicated the importance of avoiding stigmatizing effects in the design of such a public display.

**Lumilo: Co-designing Wearable Teaching Augmentation to Combine Strengths of Human and AI Instruction**

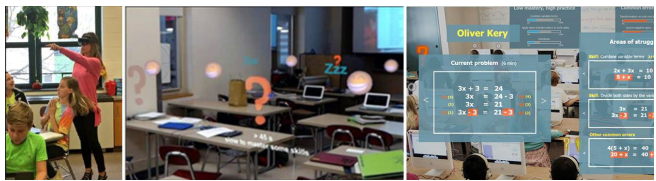

**Figure** 3. **The Lumilo project. Left: Teacher using Lumilo in class** [55]. **Middle: Real-time indicators visible at a glance. Right: Illustration of detailed screens visible on-request** [50].

The Lumilo project explored the potential for TA in AI-supported K-12 classrooms. AI-based tutoring software is increasingly used to enable more personalized instruction [48, 77, 95]. These systems allow students to work at their own pace, while adapting instruction to individual student needs. Yet AI tutors are not commonly designed to work together with human teachers to leverage their complementary strengths [51, 55]. Through an iterative series of participatory design and field studies in K-12 classrooms, Holstein et al. explored teachers' needs and desires for support during AI-supported class sessions [51, 54, 55]. One prototype that emerged is a mixed reality smart glasses application (using the Microsoft HoloLens [49]) called Lumilo [51], which augments teachers' perceptions of students' learning, metacognition, and behavior (Figure 3). When teachers glance across the classroom, they see mixed reality icons (e.g., emoji or question marks) floating above individual students' heads. These icons update in real-time based on analytics from the AI tutor. Teachers can also set ambient, spatial sound notifications in response to specific events. For example, if a student is detected as potentially misusing the software, a teacher could hear a soft sound, as if coming from that student's location in the room [51].

Lumilo is designed to alert teachers to situations that the AI may be poorly suited to handle (e.g., if the AI tutor finds its attempts to help unsuccessful, a question mark icon appears over that student's head). With such situations prioritized for teachers, they can decide when and how to intervene. In addition, Lumilo can display information upon request, to further aid teachers in deciding whether and how to help; e.g., by showing a prioritized set of areas with which a student is struggling, along with examples of specific errors the student has made (Figure 3). A field experiment showed that teachers' use of Lumilo enhanced students' learning, compared with standard AI-supported classrooms [55].

The use of smart glasses allows teachers to keep their heads up and their attention focused on the classroom, rather than buried in a screen. In early needfinding studies, teachers emphasized that much of the information they take in during a class session comes from "reading the classroom," including student body language [51]. Interfaces that would distract from signals already used by teachers, rather than augmenting them, were found to be undesirable. While this pointed to the use of spatially-distributed displays, the participating teachers were uncomfortable with information being public as in the FireFlies system, noting that some of the information most useful for their own decision-making could be stigmatizing if shared with the entire class [1, 50].

**METHODS**

We developed the TA framework by analyzing the diverse range of TA system designs reviewed above through a *constant comparative analysis* [46]. Our analysis spanned multiple bodies of literature that have introduced novel forms of teaching augmentation. Our review included work on Learning Analytics (e.g. [81, 86, 126]), AI in Education (e.g. [51, 55, 113]), CSCL (e.g. [4, 6, 32]), Teacher Education (e.g. [58, 111, 112]), and Design (e.g. [11–13]). In order to ground intermediate-level design knowledge across a wide range of interfaces, teaching contexts, and instructional goals, our analysis intentionally included cases from a wide range of existing TA systems. For instance, Figure 4 illustrates the range of interfaces included in our analysis through a non-exhaustive set of prior design cases.

During the first round of analysis, two researchers annotated prior cases to identify salient similarities and differences in their designs, while *abstracting over* specific choices of hardware interfaces or instructional contexts [46]. We then conducted *interpretation sessions* [20] to cross-check annotations for individual cases, and to synthesize a set of design dimensions that captured variation across existing TA systems. For example, during this phase we found that prior design cases clustered based on the extent of *pre-interpretation* a system performed before presenting information to the teacher (see D5 in the next section). At one extreme, some systems were designed to present low-level, minimally processed information, leaving more space for teachers to apply their own interpretations. At the other, some systems were designed to perform substantial upfront interpretation on behalf of the teacher, scaffolding teachers towards particular understandings of ongoing events.

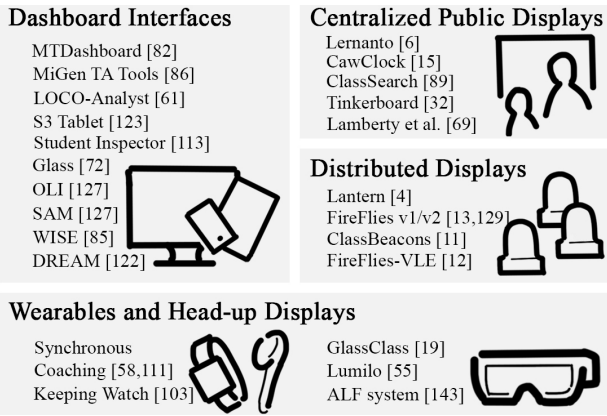

Figure 4. Examples of design cases included in our analysis.

Using these emergent dimensions, we re-analyzed the design cases in a second round of analysis. This re-analysis enabled us to refine our dimensions in order to better capture design variation across different systems. In the next section, we present the design dimensions that emerged from our analysis and contextualize them using concrete examples from existing TA designs.

## TEACHING AUGMENTATION DESIGN FRAMEWORK

In this section, we present the TA (Teaching Augmentation) framework, consisting of five dimensions that reveal a rich design space for TA systems. We examine and contextualize each dimension of the TA framework through existing design cases – discussing how points along each dimension are instantiated by features in existing systems. In doing so, we surface key design decisions and trade-offs, which may otherwise remain implicit in designed artifacts.

### D1: Target — Which abilities are being augmented?

The first dimension in the TA framework addresses which teacher abilities are being augmented. Real-time teaching augmentation can have various goals, including augmenting what teachers *notice* during ongoing instruction, what situations they *act upon*, and what *actions* they take.

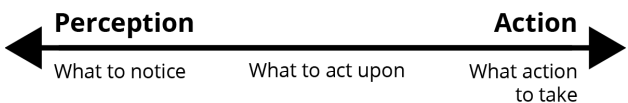

Figure 5. The spectrum of D1: Augmentation Target.

*What to notice*: Teacher awareness tools [53, 83, 107] are commonly designed to augment teachers' perceptions of what goes on in their classrooms. For example, the ClassBeacons system extends a teacher's ability to monitor critical yet implicit aspect of their own ongoing behavior: their distribution of proximity across students while circulating around the classroom [10]. Similarly, systems like Lantern [4] and MTDashboard [82] extend a teacher's ability to monitor students' progress or performance during ongoing classroom activities.

*What to act upon*: In addition to supporting teachers' situational awareness, TA systems may also help teachers in efficiently *prioritizing* situations that require them to act. For example, the MTFeedback system [81] actively directs teachers towards particular collaborating groups who may be most in need of a teacher's help, based on real-time analytics about groups' performance and collaboration quality [81]. Similarly, Lantern supports teachers in prioritizing help among students by showing help requests from student groups together with the amount of time a group has spent waiting [4]. The Lumilo system directs teachers' attention towards situations that AI tutors may be poorly suited to handle [51, 55], while teachers decide whether, when, and how to help particular students.

*What action to take*: Finally, at the right end of this dimension, technologies may be designed to support teachers in deciding *what action to take* in particular situations, or even to *automate certain teacher tasks*. For example, "synchronous coaching" approaches [58, 111] use a wireless earpiece to give (novice) teachers live advice about how to handle unfolding situations during a class session. This advice is provided by a more experienced coach who remotely monitors the session. Beyond teacher recommendations, intelligent learning technologies such as AI tutoring systems [52, 54, 104, 105, 141] may help the teacher automate some of their tasks, scaling the teacher's capacity to act or freeing up their time for other activities.

**Tension: Autonomy versus Automation.** Analysis of the different design cases suggests a delicate tension between teacher needs for automation versus autonomy in the classroom. On the one hand, over-automation risks taking over classroom roles that teachers would prefer to perform and may limit their flexibility to set their own instructional goals [10, 47, 52], or restrict their abilities to productively *improvise* [117]. Yet under-automation risks burdening teachers with tasks they would rather not perform (e.g., routine grading), limiting their time available for other activities. It may also restrict the extent to which they can feasibly tailor instruction to individual student needs [51, 52, 94, 125]. Given these trade-offs, it is critical to understand the roles teachers prefer to perform in particular educational contexts, and where automation may help more than hurt [28, 47, 52, 76]. A central challenge in the design of teaching augmentation is to find balances of teacher autonomy and automation that serve both teacher and student needs [47, 52].

**D2: Attention — Which attentional levels are targeted?**
The second dimension of the TA framework asks which attentional level(s) are being targeted. TA systems could augment teachers' practice at various levels along their *attentional continuum* [16]: from the central (focal) area of their attention all the way to their periphery of attention.

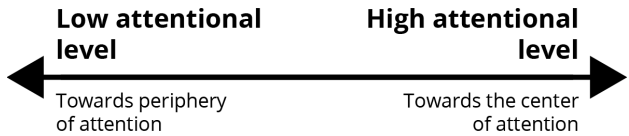

Figure 6. The spectrum of D2: Attention.

*Towards the periphery:* Real-time teaching augmentation may be designed to require little focal attention from the teacher, leveraging the periphery of their attention [14, 16]. For example, classroom light objects such as Lantern [4] and FireFlies [129], provide peripheral, spatially-distributed information at a low-resolution (e.g., via visual signals such as color, intensity, and blink rate). Lumilo offers glanceable information via icons above students. Systems such as CawClock [15] additionally utilize soundscapes to augment teachers' time awareness in an unobtrusive manner.

*Towards the center of attention:* Towards the other end of the spectrum, TA systems may be designed to require continuous focal attention from the teacher. For example, most mobile or tablet-based learning analytics dashboards present rich data visualizations in a centralized fashion [2, 25, 83, 127]—requiring teachers to switch their focus between the dashboard and the ongoing classroom activities [4, 10, 51]. Similarly, approaches such as synchronous coaching [58, 111] require a teacher to focus, at least temporarily, on verbal advice from a human coach [58].

**Tension: Unobtrusiveness versus Informativeness.** This dimension is critical to consider given how scarce a teacher's attention can be during an ongoing class session [9, 50], as also demonstrated in research on *orchestration load* [101, 102]. Yet at present, the attentional continuum [14, 16] is rarely an explicit consideration in the design of teacher-facing classroom technologies [9, 10, 34]. To date, relatively few projects have explored the design of peripheral TA features (e.g., [4, 15]). Peripheral modes promise to keep teachers' attention focused on the classroom, acknowledging the complexity of their ongoing task performance, and helping teachers remain attentive to classroom signals that may not be captured in a system's data streams (e.g., student body language) [10, 51]. However, the information capacity of a teacher's periphery is limited. At moments, teachers may wish to engage with the information at TA system provides at a more detailed level than is possible without a focal display. A core challenge for the design of TA systems is to effectively integrate focal and peripheral modes of display – providing information at the periphery of teachers' attention when possible, but allowing teachers to effortlessly access focal displays when deeper engagement is needed.

**D3: Social Visibility — With whom is awareness shared?**
The third dimension of the TA framework asks which information is visible to which classroom stakeholders.

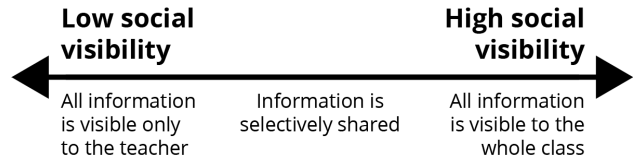

Figure 7. The spectrum of D3: Social Visibility

*Low social visibility:* Since real-time teaching augmentation is primarily aimed for enhancing teachers' perception or cognition, TA systems may be designed to provide information that is only visible to the teacher. For example, Lumilo's smart glasses interface (or other wearable systems [19, 58, 103, 143]) ensures that the information provided to teachers remains private from students or others present in the classroom.

*High social visibility:* At the other end of the spectrum, information may be socially visible to everyone in the classroom, as in the Lantern and FireFlies systems [4, 129]. For example, in classrooms using Lantern, students can simply look around the classroom to see information about other students' progress and current help needs [4]. ClassBeacons affords shared accountability for teachers and students to influence teacher proximity, which is otherwise often subconsciously delivered and passively received [11].

*Selective shared awareness:* In between these two ends of the spectrum lies a rich design space for systems that facilitate real-time teacher control over *which information* is shared with *which individuals or groups* [50, 52, 82]. For example, in addition to enabling teachers to privately monitor student work during an ongoing class session, the MTDashboard system allows teachers to selectively share what they are seeing with the whole class by "sending" examples of student work to a projected wall display [82].

**Tension: Privacy versus Mutual Accountability.** Prior design research reveals teacher and student preferences for certain kinds of information to be kept private during a class session [12, 51, 52]. For example, information that could be perceived as negatively assessing individual students should generally not be shared with the whole class [51, 52]. At the same time, teachers and students have expressed needs for some level of shared visibility during class, in order to promote mutual accountability [11, 33], to help motivate students through competition or cooperation [4, 52] or to facilitate students to support peers in need of help [52, 132]. In light of these teacher and student needs, a central challenge in the design of TA systems is to anticipate student and teacher boundaries regarding the social visibility of real-time information in the classroom, and to design forms of shared awareness [91] that can effectively support student motivation and collaboration without crossing these boundaries [52, 93].

**D4: Presence over Time — When is TA available?**
The framework's fourth dimension asks how continuously the augmentation is present during an ongoing class session.

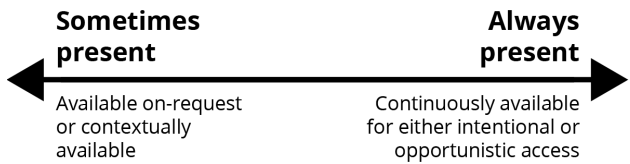

Figure 8. The spectrum of D4: Presence over Time

*Always present*: Real-time teaching augmentation can be designed to provide information continuously, so that it can be perceived at any time without requiring the teacher to intentionally access it. For example, using TinkerBoard [32], while helping a particular learner team, a teacher can gain basic awareness about the other teams at any time. Similarly, ClassBeacons supports opportunistic sensemaking [10]: in the midst of ongoing tasks, a teacher may *happen to notice* that he or she has spent too little time in a particular region of the classroom, potentially prompting reflection.

*Sometimes present*: Alternatively, or in addition to providing information continuously, TA systems may provide certain information only upon a teacher's request or in specific situations. For example, the technology probe presented in [103] provides haptic notifications upon classroom events deemed currently relevant. As another example, although the Lumilo glasses present glanceable icons at all times – floating above individual students' heads to indicate their current 'states' – this system prioritizes which icon is immediately visible to the teacher. If Lumilo detects multiple states at once for a given student (e.g., if a student is both *unproductively struggling* and *avoiding using hints*), only one icon will be shown at a glance, based on a priority queue [51]. Other currently active states are provided only upon a teacher's request, by clicking on the student's icon.

**Tension: Opportunistic Sensemaking versus Prioritizing.** TA systems that present information continuously hold the potential to support opportunistic sensemaking and to provide teachers with greater agency over how and when to deliberate on relevant but not time-critical phenomena (e.g., they may perceive the information and reflect on it in moments where they are less busy). At the same time, by presenting information more selectively, designs towards the left end of this dimension may better support teachers in prioritizing classroom situations that most require their attention or intervention in-the-moment. Thus, a central challenge in the design of TA systems is to support the opportunistic sensemaking of teachers, while simultaneously ensuring that such opportunism does not distract teachers from noticing and addressing critical phenomena in the classroom.

**D5: Interpretation — How is the task of interpretation shared between the teacher and the system?**
Finally, the framework's fifth dimension asks how the task of interpreting classroom situations is divided between the teacher and the system. No TA system can be truly agnostic [33, 45, 67]: no matter how simple and direct an information display may be, its design will always represent a particular framing of ongoing classroom events (e.g., by the mere act of including certain kinds of information while excluding others). Nonetheless, TA systems may be designed to perform relatively more or less *pre-interpretation* of information before presenting it to teachers.

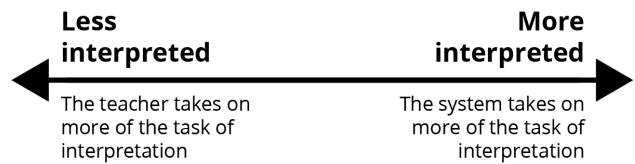

Figure 9. The spectrum of D5: Interpretation.

*Less interpreted:* Towards the left end of this dimension, real-time teaching augmentation may be designed to present lower-level, less-processed information, leaving more room for teachers to apply their own interpretations. For example, ClassBeacons visualizes a teacher's distribution of proximity across students, but without interpreting whether a distribution is "good" or "bad" [11]. Similarly, many existing learning analytics dashboards present teachers with low-level metrics such as the cumulative amount of time a student has spent on an assignment [51, 107], or minimally-interpreted data portrayals of learners' physical actions (see the case of Collaid [127] as an example).

*More interpreted:* Towards the other end, real-time teaching augmentation may be designed to present teachers with higher-level, more immediately meaningful information, by having the system provide more upfront interpretation on behalf of the teacher [33, 54, 109] For example, MTDashboard provides a layer of interpretation by visualizing the distance between a student's work (on a concept mapping task) and an expert-generated solution [82]. Similarly, systems like FACT [125] and Lumilo use student modeling methods [29] to provide teachers with real-time indicators of higher-level student constructs, such as whether a student is making *effective use of help* [108].

**Tension: Providing Higher-level Interpretation versus Privileging Teachers' On-the-ground Knowledge.** By leaving more interpretation up to the teacher, designs toward the left end of D5 have the potential to facilitate more accurate and meaningful teacher inferences – taking advantage of the rich contextual knowledge to which teachers have access but the system may not (e.g., prior knowledge about individual students, or relevant classroom events not captured in the system's data streams) [5, 10,

50]. Leaving space for interpretation can also encourage teachers to make use of the information based on their own didactic styles or classroom cultures. By contrast, designs toward the right end of D5 risk imposing inaccurate or contextually-irrelevant judgments on the teacher [5, 10], or suggesting actions that do not match their didactic styles. At the same time, given limited time and attention during ongoing instruction, teachers may need more scaffolding from the system in deriving actionable insights from data in-the-moment. Prior design research suggests that teachers sometimes find the kinds of information commonly presented in learning analytics dashboards to be too low-level to be useful [54, 107]. For example, beyond simply seeing how frequently a student is making errors on a learning task, teachers might want support in interpreting whether the student truly needs their help (e.g., whether or not this student is struggling *productively*) [51, 54, 62]. Given these trade-offs, a core design challenge for TA systems is to find the right "blend" of human and machine intelligence for specific educational contexts [5, 47, 51, 94].

**DISCUSSION**

Based on the design dimensions and trade-offs explicated in the TA framework, we first discuss implications and opportunities for future design. We then demonstrate how the TA framework can help designers and researchers characterize and compare the patterns of prior TA solutions.

**Implications and Opportunities for Future Design**

*Design to Balance Automation with Teacher Autonomy*

As discussed above, a central tension in the design of any teaching augmentation system is the balance between *autonomy* and *automation* (Dimension D1). To help designers find effective balances, we offer three guiding questions for designers to consider in the early stages of a project (e.g., context and needfinding studies). First, which specific teaching tasks will be augmented? Teachers' needs for professional autonomy may differ substantially across different types of tasks (e.g., assessing student work, providing emotional support, assessing and managing student motivation) [51, 52, 76], yet such preferences can be quite nuanced and may run counter to designers' initial intuitions [28, 52, 94]. Second, what do teachers see as their role(s)? Teachers' needs for autonomy may depend on how they understand their own roles in a given context. For example, teachers working in lecture-heavy contexts may perceive very different needs than teachers who act more as facilitators in self-paced lessons [10, 12, 54, 125, 140, 143]. Teachers teaching different age groups of students may also have different views of their roles and needs for autonomy. Finally, how experienced are the targeted teachers? Forms of augmentation that less experienced teachers perceive as helpful may be perceived as intrusive by teachers with greater experience and skill [50, 64].

*Design for Seamless Transitions on the Attention Continuum*

Dimension D2 represents a core tension between *unobtrusiveness versus informativeness* in the augmentation a TA system provides. While the information presented by TA systems needs to be informative enough to benefit teachers' practice, it also needs to be unobtrusive enough for use *during* their busy ongoing activities, without overburdening their attention. The key to addressing this challenge is to design information displays that can meaningfully inform teachers in different *attentional levels* [16] —enabling seamless transitions between the periphery and the center of their attention. To this end, we propose two concrete design opportunities for future TA systems. The first opportunity is to combine focal interfaces with peripheral design features. Focal interfaces (e.g., dashboard visualizations) have rarely been designed to include peripheral information representations, *in addition* to centralized, more detailed representations. Integrating such peripheral features could reduce teachers' attentional threshold for using these focal interfaces, extending their usefulness during ongoing action (e.g., they can continue to provide value even as teachers are absorbed in other activities, such as helping students). The second opportunity is to combine visual modalities with secondary channels. Studies on human attention [87, 110, 139] suggest that information streams reaching different sensory channels (e.g., reading while listening to music) are less likely to cause attentional conflicts than those reaching the same channel (e.g., listening to two songs). Therefore, distributing supportive information over *multiple* sensory channels could further enable teachers' unobtrusive perception of the information [9, 17]. While vision has been predominantly used among existing designs, a few cases also suggest sound as a promising assistive channel to convey information both continuously (e.g., soundscape [15]) and discretely (e.g., ambient sound notifications [51] and live coaching via an earpiece [58]). A fruitful direction for future design is to further explore how auditory or haptic modalities could be used as alternative or additional channels to seamlessly augment teachers' practice.

*Design to Balance Benefits and Risks of Shared Awareness*

Dimension D3 represents a design tension between *privacy* and *mutual accountability*, echoing the theory of social translucence [36] in the specific case of classroom pedagogy. TA systems that afford shared awareness [91] hold the potential to facilitate collaboration [4, 52, 131] and mutual accountability [10, 11] among stakeholders in the classroom. However, in some contexts, shared awareness introduces the risk of undesirable social effects (e.g., unhealthy peer comparisons or stigmatization) [1, 60]. To navigate this tension, beyond simply considering *whether* to share awareness with students, designers should consider *how to present* information to different stakeholders in the classroom (teachers versus students), *which students* should be able to see *which information*, and under *which circumstances*. As identified in the TA framework, there are rich design possibilities in the middle region of D3 to explore in future work, for example: (1) Exploring different levels of *anonymity* or *ambiguity* in information shared beyond the teacher; (2) Understanding social boundaries (among both students and teachers) regarding which kinds of information can be shared [52]; and (3) Exploring who should have the control over what and when to share (e.g., which classroom stakeholders get to judge the appropriateness of sharing in particular situations [52, 81]).

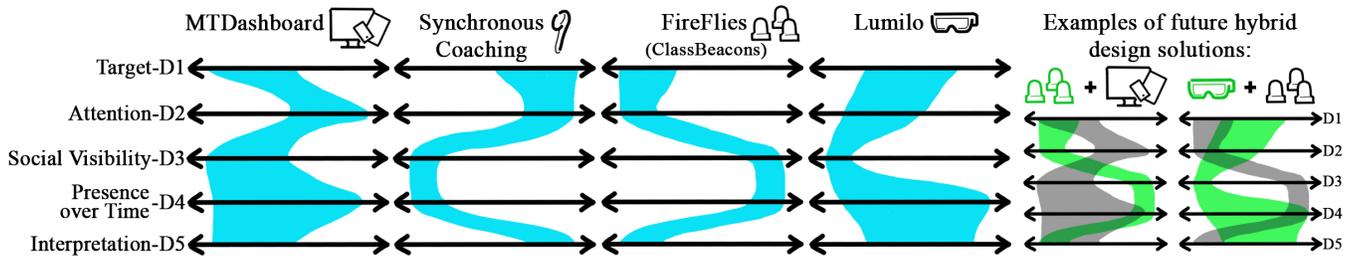

Figure 10. Examples of how the TA framework can be used to characterize the patterns of existing design cases, and point towards a broad design space for *hybrid* design solutions (of which just two examples are shown) that combine advantages of prior cases.

*Design to Balance Opportunism with Prioritization*
Dimension D4 surfaces the design tension between *opportunistic sensemaking* and *prioritization*. Continuously presented information holds the potential to leverage teachers' opportunistic sensemaking, enabling unplanned, serendipitous reflections on pedagogically relevant information during class [10]. On the other hand, information presented at discrete points in time may better support teachers in prioritizing those situations that most require their attention in-the-moment. Therefore, one relevant consideration in navigating this tension is the *time-sensitivity* of the presented information. Along this dimension, we identify two fruitful opportunities for future design. First, future work should explore how to design *information-on-request* in peripheral displays. Current peripheral displays (e.g., Lernanto [7] or FireFlies [12]) are typically designed to continuously present the same set of information. Future designs could support teachers in toggling between various types of information based on which information is considered most relevant in the current classroom activity. Second, future work should explore *activity-aware* [70] features for information presented at discrete points in time. While rarely explored in prior TA designs, activity-awareness can support the prioritization of information to present, based on teachers' current task context. E.g., when a teacher shifts from whole class lecturing to helping a student-team with a specific topic, a TA system might automatically present information about these students' prior difficulties with this topic.

*Design to Combine Strengths of Human and Machine Interpretation*
Dimension D5 represents a tension between *privileging teachers' on-the-ground knowledge* versus *providing high-level interpretations* [5]. Underlying this tension is the question of how to optimally combine the strengths of teacher and machine interpretations. Below, we offer three broad recommendations for future design. First, designers should intentionally "leave room" for teacher interpretation in cases where meaning is likely to be highly context dependent, yet an automated system is unlikely to have all of the necessary context [5, 10, 51]. For example, TA systems might leave room by providing *minimal system interpretation* [10, 30] *expressing appropriate uncertainty* about their own interpretations [51] or otherwise *prompting teachers to verify* or *override* their interpretations. Second, beyond *leaving room* for teacher interpretation, designers should support teachers in *productively* second guessing a TA system's interpretations. For example, a system that presents high-level interpretations should also offer insight into how its interpretations were made [74, 134], or enable teachers to view relevant, less-interpreted data, to check alignment between the system's interpretation and their own [5, 51, 74]. Finally, designers should consider ways to allow teachers to *meaningfully participate* in shaping a TA system's interpretations [68, 106]. This might be achieved by enabling 'conversations' between the teacher and the system [23, 51]—for example, by allowing teachers to provide feedback on system interpretations, or by having the system proactively request teachers' input regarding contextual information to which the teacher has unique access (e.g., via an interface similar to Group Spinner [66]).

**Studying the Patterns of Existing TA Design Solutions**
The TA framework presents a broad framing of teaching augmentation (TA) to facilitate the analysis and design of a diverse range of systems, developed across multiple research communities. Here we demonstrate the utility of the framework by exploring patterns (cf. [42]) of several existing TA system designs: MTDashboard [82], synchronized coaching systems [58], FireFlies [10], and Lumilo [55]. Through the lens of the TA framework, these cases can be visualized as shown in Figure 10.

This visualization surfaces underlying commonalities and differences across designs. For example, the ClassBeacons implementation of FireFlies [10] and the synchronous coaching system [58] have nearly mirrored patterns (Figure 10), covering inverse regions on all dimensions. This is due to differences in these systems' design goals. By providing real-time advice, the synchronous coaching system aims to enable novice teachers to develop effective repertoires for dealing with various classroom situations [22, 58]. In contrast, by prompting teachers to reflect in action,

FireFlies aims to help teachers optimize their established routines, and avoid rigidity in their practice [10, 116, 117]. As a result, the synchronous coaching system *prescribes teacher action* (D1) in response to classroom moments *deemed critical* (D4) based on the interpretations of an *expert coach* (D5), whereas FireFlies *extends teacher perception* (D1) to promote *opportunistic sensemaking* (D4) relying on *teachers' own interpretations* of the data (D5). Moreover, the synchronous coaching system offers voice instructions (D2) via a *private channel* (D3), while FireFies uses public ambient signals (D2), aimed for also increasing *students' accountability* for teacher proximity (D3).

The pattern visualization also reflects the difference between more broadly versus more narrowly focused TA systems. More broadly focused TA systems like MTDashboard and Lumilo often contain multiple layers or channels of information representation. As a result, they tend to cover a larger area across the five dimensions (Figure 10). By contrast, more narrowly focused TA systems, like FireFlies or the synchronous coaching system, often contain less layers or channels, and thus tend to cover a smaller area. Such differences can arise due to differences in these systems' design goals, or inherent affordances of the modalities they use (e.g., screens versus ambient lamps or earphones). However, this does not imply that TA systems covering a larger pattern area are necessarily better, or that they can replace those with a smaller area. On the contrary, these existing solutions have the potential to complement each other in many ways, as discussed below.

*Opportunities in the Design of Hybrid Solutions: Towards TA Systems as Interface Ecologies*

As shown in Figure 10, a set of distributed peripheral displays (e.g., FireFlies) could be meaningfully combined with a (tablet) screen-based dashboard interface (e.g., MTDashboard). The distributed lamps could provide a layer of peripheral representation that enables teachers to tacitly perceive information (e.g., the state of each student group) with a low attentional threshold and minimal interruption to their ongoing tasks. Meanwhile, if teachers notice relevant information via this peripheral layer (e.g., several groups are currently struggling with their tasks), they could check the dashboard interface to access more detailed information (e.g., details about the nature of a group's difficulties). Moreover, the distributed lamps can create an ambient channel for teachers and students to communicate non-verbal information or share awareness about classroom activities, which may facilitate shared orchestration and collaboration in the classroom. Similarly, as shown in Figure 10, a pair of smart glasses for teachers (e.g., Lumilo) could be meaningfully combined with a distributed display (e.g., FireFlies): while the distributed display enables shared awareness and accountability among both teachers and students, the smart glasses can display high-level constructs about individual learners as extra information visible only to teachers. The glasses could actively steer teachers' attention to actionable situations without interrupting or stigmatizing students.

Although rarely explored in the literature, the above examples illustrate potential for such hybrid solutions. Similarly, many more hybrids are possible (e.g.[58]+[103]). We envision that in the future, increasingly ubiquitous and interconnected classroom technologies will serve as an interface ecology [21] that seamlessly augments teachers' surroundings. Thereby, future TA systems can ultimately be designed across devices, to support teachers at the right moments, in the right places, and with seamless integration into their ongoing (cognitive and physical) tasks.

## CONCLUSIONS AND FUTURE WORK

We have introduced the TA framework, which aims to inform the design and analysis of teaching augmentation systems by: (1) revealing underlying, often implicit design considerations with relevance across many kinds of TA systems, and (2) proposing a common lens through which researchers in this interdisciplinary area can analyze diverse existing design cases and jointly generate new insights.

Our generalization of the TA framework is not meant to be exhaustive or definitive. For example, sub-dimensions might be derived from D4 to respectively address the temporal and spatial relevance of information provided by TA systems. We encourage researchers and designers to enrich, modify, or challenge our generalization through design practice, in order to generate deeper understandings for the field. An exciting direction for future work is to explore the extent to which the TA framework generalizes to a broader range of educational contexts, including special education contexts and informal learning settings (e.g., museums) in which an instructor is present. More broadly, we expect that many of the insights reflected in the TA framework may generalize to a broader range of "caring professions," (such as social work and nursing) where real-time augmentation can seamlessly empower human practitioners, but must avoid obtruding or over-automating [24, 115, 142].

In sum, the TA framework is intended to help researchers and designers more comprehensively explore the possibility space for teaching augmentation systems. It is our hope that this framework will support designers in combining insights from existing TA designs that have emerged across different domains, and to analyze underlying trade-offs in their own design decisions.

## ACKNOWLEDGEMENTS
We thank our anonymous reviewers, as well as our colleagues Dr. Rong-Hao Liang and Dr. Javed Khan, for their valuable feedback during our writing.